# Understanding Large-Scale Plasma Simulation Challenges for Fusion Energy on Supercomputers

J. J. Williams[1], A. Bhole[1], D. Kierans[1], M. Hoelzl[2], I. Holod[2], W. Tang[2], D. Tskhakaya[3], S. Costea[4], L. Kos[4], A. Podolnik[3], J. Hromadka[3], JOREK Team[5], E. Laure[6] & S. Markidis[1]

[1] *KTH Royal Institute of Technology, Stockholm, Sweden*
[2] *Max Planck Institute for Plasma Physics, Garching and Greifswald, Germany*
[3] *Institute of Plasma Physics of the CAS, Prague, Czech Republic*
[4] *LeCAD, University of Ljubljana, Ljubljana, Slovenia*
[5] *See the author list of [Nuclear Fusion, Doi: 10.1088/1741-4326/ad5a21]*
[6] *Max Planck Computing and Data Facility, Garching and Greifswald, Germany*

*jjwil@kth.se*

**Introduction**

Understanding plasma instabilities is crucial for sustainable fusion energy. Large-scale plasma simulations are essential for designing next-generation fusion energy devices and modeling industrial plasmas. Accurate modeling and prediction of plasma behavior under extreme conditions require sophisticated simulation codes to capture the complex interactions between plasma dynamics, magnetic fields, and material surfaces. This work conducts an HPC analysis of two prominent plasma simulation codes, BIT1 and JOREK, to advance the understanding of plasma behavior in fusion energy applications. The focus is on evaluating JOREK's computational efficiency and scalability for simulating non-linear MHD phenomena in tokamak fusion devices [1]. Previous studies examined BIT1 [2, 3], a massively parallel PIC code for fusion device plasma-material interactions. Investigations into BIT1's computational requirements and scalability on supercomputers provided crucial insights. Detailed profiling pinpointed bottlenecks, guiding optimization efforts that notably boosted parallel performance. Key contributions include addressing challenges in BIT1's neutral particle ionization and non-linear JOREK shattered pellet injection (SPI) simulations. Profiling techniques also highlighted computational hotspots during strong scaling tests, especially with activated diagnostics.

**PIC MC BIT1 Code**

Modeling plasma-loaded divertors in fusion devices like ITER is challenging. The divertor manages heat and particle fluxes, protects the first wall from high-energy neutron damage, and removes impurities by diverting plasma flow to a specific region at the bottom of the toroidal chamber (see Fig. 1). The Particle-in-Cell (PIC) method models plasma behavior by simulating particle dynamics in 1-3D spatial dimensions and 3D velocity space, integrating Monte Carlo (MC) routines for particle collisions and interactions with chamber walls. The PIC cycle involves interpolating particle data for plasma density computation, smoothing densities to remove spurious frequencies, solving linear systems for electric and magnetic fields, using MC techniques for collisions and wall interactions, and advancing particle positions and velocities. BIT1, a 1D3V PIC code optimized for HPC systems, operates in one-dimensional space with three velocity components per particle, facilitating large-scale plasma simulations.

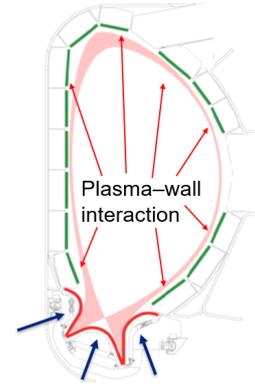

Figure 1: *BIT1 simulates plasma behavior in the tokamak divertor (blue arrows), such as in the ITER fusion device [3–5].*

**Non-linear Extended MHD JOREK Code**

JOREK, an HPC-capable code for fusion reactor modeling [1], employs hybrid MPI+OpenMP parallelization and various plasma physics models, including fluid and kinetic ones. It uses advanced finite element methods and fully implicit time-integration techniques with GMRES for solving linearized equations. Interfaces to libraries like PaStiX and STRUMPACK enhance GMRES efficiency. Given its extensive equations involved, simulations of fusion reactors are computationally demanding on supercomputers due to their scale in length and time. As an example, in the Fig. 2, a snapshot of the toroidal current is plotted at 0.72 ms in a 3D SPI simulation performed using two-temperature reduced magentohydrodynamics (MHD) model [1]. The simulation assumes a frozen pellet of Neon gas being shattered into 53 fragments that are injected into ASDEX-upgrade plasma. It uses 12,460 finite elements in the poloidal plane and up to 7 toroidal harmonics resolved using 64 equidistance planes.

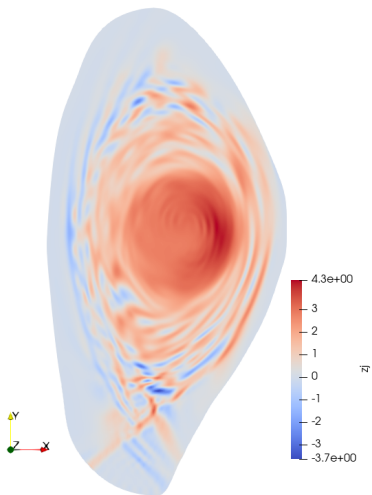

Figure 2: *SPI in ASDEX-upgrade simulation using JOREK performed on Marconi: a snapshot of the toroidal current ($j_\phi$) plotted in a poloidal plane at 0.72 ms.*

**Methodology & Experimental Setup**

In this work, we use `perf` for profiling hardware performance counters, focusing on execution time, cache, and memory. Additionally, we employ `TAU` (Tuning and Analysis Utilities) for automated and manual instrumentation to analyze performance data, identifying bottlenecks and inefficient code paths [6]. Our experiments compare BIT1 and JOREK on two systems: `Marconi A3 (Skylake)`, featuring 2982 nodes with Intel Xeon 8160 24-core processors and `Dardel`, with 1270 nodes equipped with AMD EPYC Zen2 64-core processors. Further details about these systems are provided in the poster.

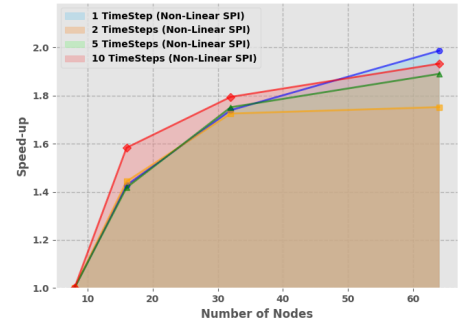

Figure 3: *JOREK SPI simulation speed-up on Marconi up to 64 nodes.*

**Performance Results and Analysis**

We begin by identifying the challenges discovered in the BIT1 neutral particle ionization and non-linear JOREK SPI simulations to pinpoint the most computationally intensive parts of these codes during strong scaling tests, particularly with diagnostics activated. As previously discovered and improved in [2, 3], using `perf`, the BIT1 performance considerably depends on the problem size and effective LLC (L3) usage. When BIT1 simulates up to 50 nodes on Dardel in [2], the Ionization simulation scales well initially up to 10 nodes but then faces diminishing returns with reduced efficiency as more nodes are added beyond 20. This highlights the challenges in maintaining linear speedup with increasing parallelism for small size problems. Contrary to this, large size BIT1 simulations indicate hyper-scaling [2]. For JOREK, we analyze a 3D SPI simulation from a checkpoint to study JOREK's HPC performance. In Fig 3, we compile JOREK with STRUMPACK v7.2.0 and compare 8 to 64 nodes on Marconi. Significant speed-ups are observed using `perf`: for 10 time steps, speed-up increases from 1 (8 nodes) to 1.93 (64 nodes), a 93% improvement; for 5 time steps, it goes from 1 to 1.89, an 89% gain. Fig 4 shows 10 time steps with varying OpenMP threads on Marconi, revealing decreasing execution time with more threads, but not at an ideal linear rate. Using 2 threads as a baseline (3532.7 seconds), 4 threads take 2235.1 seconds (26.6% slower than ideal). Increasing to 8, 12, and 24 threads results in 1543.7, 1389.6, and 1377.3 seconds, respectively, versus ideal times of 883.2, 588.8, and 294.4 seconds. This shows diminishing returns due to overhead from thread management and synchronization.

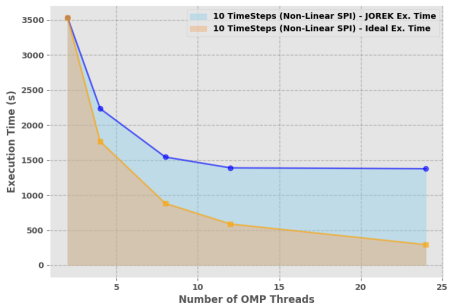

Figure 4: *JOREK SPI simulation for 10 time steps on 16 nodes up to 24 OpenMP threads on Marconi.*

Using `TAU` profiling on JOREK SPI simulations for 10 time steps on `Marconi` reveal detailed performance insights. Simulating with 2 OpenMP threads, `thread 0` pri-marily focuses on *construct_matrix* (58.1%) and *solve_sparse_system* (32.6%), while `worker threads` are fully occupied by executing the *OpenMP_Thread_Type_ompt_thread_worker* routine (contributing 39.1% each to *elementary_matrix_build* and *element_matrix_fft*). Increasing to 24 threads shifts `thread 0`'s focus to *solve_sparse_system* (53.2%) and *construct_matrix* (35%), optimizing sparse system solving with enhanced parallel capabilities. As seen in Fig 5, `worker threads` maintains focus on the *OpenMP_Thread_Type_ompt_thread_worker* routine, however it contributions to *elementary_matrix_build* and *element_matrix_fft* reduction by 10.6% each, highlighting improved workload distribution across threads for efficient performance in complex computations.

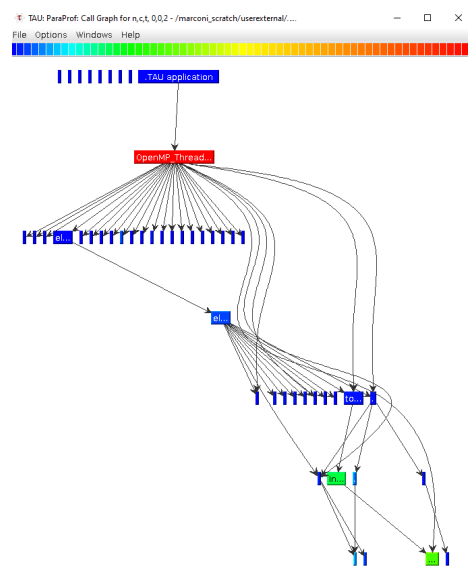

Figure 5: *TAU profiling worker threads call graph with OpenMP routine (in red) on Marconi.*

**Conclusions and Future Work**

In this work, we identified the *solve_sparse_system* and *construct_matrix* routines as the most computationally intensive parts of the JOREK SPI simulation, consuming up to 88.2% of execution time with 24 threads. This analysis was conducted for a moderate-size problem, and better scalability is expected for the largest production cases. Future efforts will extend our analysis to other platforms and incorporate hybrid fluids+kinetic models to enhance overall efficiency.

**References**

[1] M. Hoelzl, et al., NF **61**, 065001 (2021)
[2] J. J. Williams, et al., LNCS **14351**, Springer Nature, arXiv:2306.16512 (2023)
[3] J. J. Williams, et al., LNCS **14832**, Springer Nature, arXiv:2404.10270 (2024)
[4] D. Tskhakaya, et al., Contrib. to Plasma Phys., volume **44**, 5-6, 564-570 (2004)
[5] D. Tskhakaya, et al., Pisa Italy 2010, Computer Society, 476-481, IEEE (2010)
[6] Shende, Sameer S., et al., volume **20.2**, 287-311 Sage Pub. Sage CA (2006)
[7] Snyder, Shane, et al., 5th Workshop on ESPT, 9-17, IEEE (2016)

**Acknowledgements**

*This work has received funding from the European High Performance Computing Joint Undertaking (JU) and Sweden, Finland, Germany, Greece, France, Slovenia, Spain, and the Czech Republic under grant agreement No 101093261. The computations/data handling were/was enabled by resources provided by the National Academic Infrastructure for Supercomputing in Sweden (NAISS), partially funded by the Swedish Research Council through grant agreement no. 2022-06725. Parts of this work was carried out using the Marconi-Fusion Supercomputer.*